\documentclass[12pt,preprint]{aastex}
\def\msol   {\hbox{$M_\odot$}}                  
\def\ddeg   {\hbox{$.\!\!^\circ$}}              

\begin{document}

\title{Flaring Activity of Sgr A* at 43 and 22 GHz: Evidence for
 Expanding Hot Plasma}

\author{F. Yusef-Zadeh\footnote{Department of Physics and Astronomy,
Northwestern University, Evanston, Il. 60208
(zadeh@northwestern.edu)},
D. Roberts\footnote{Adler Planetarium and Astronomy Museum, 1300
South Lake Shore Drive, Chicago, IL 60605
(doug-roberts@northwestern.edu)},
M. Wardle\footnote{
Department of Physics, Macquarie University, Sydney NSW 2109,
Australia (wardle@physics.mq.edu.au)},
C. O. Heinke\footnote{Department of Physics and
Astronomy,
Northwestern University, Evanston, Il. 60208
(cheinke@northwestern.edu)} and 
G. C. Bower\footnote{Radio Astronomy Lab, 601 Campbell Hall,
University of California, Berkeley, CA 94720
(gbower@astron.berkeley.edu)}}

\begin{abstract}

We have carried out Very Large Array (VLA) continuum observations to study 
the variability  of Sgr A* 
at 43 GHz ($\lambda$=7mm) and 22 GHz ($\lambda$=13mm). 
A low level of flare activity has been detected with a duration of $\sim$ 2 
hours at 
these frequencies, showing the peak flare emission at 43 GHz  leading 
 the 
22 GHz peak flare by  $\sim20$ to 40 minutes.  The overall 
characteristics of the flare 
emission are interpreted in terms of the plasmon  model of Van der Laan 
(1966) by considering
the ejection and adiabatically expansion  of a uniform, spherical plasma blob 
due to flare activity.
The observed peak of the  flare emission with a 
spectral index $\nu^{-\alpha}$ of $\alpha$=1.6 is
 consistent with the prediction that the
peak emission shifts toward lower frequencies in an 
adiabatically-expanding  self-absorbed source.  
We 
present the expected synchrotron light curves for an expanding blob as 
well as the peak frequency emission as a function of the energy spectral 
index constrained by the available flaring measurements in near-IR, 
sub-millimeter, millimeter and radio wavelengths. We note that the blob 
model 
is consistent with the available measurements, however, we can not rule 
out the jet of Sgr A*. 
If 
expanding material leaves the gravitational potential of Sgr A*, the total 
mass-loss rate of nonthermal and thermal particles is estimated to be 
$\le 2\times10^{-8}$\,M$_\odot$\,yr$^{-1}$. We discuss the implication
of the mass-loss rate  since this value matches
closely with the estimated accretion rate based on 
polarization measurements.

\end{abstract}


\section{Introduction}

There is now compelling evidence that the compact nonthermal radio source 
Sgr A$^*$ is identified with a massive black hole at the center of the 
Galaxy. Stellar orbit measurements have shown a mass 3-4 $\times 10^6$ 
\msol\ coincident within 45 AU of Sgr A*, and have created a much better
picture of this object since its discovery more than 
30 years ago (Balick \& Brown 
1974; Sch\"odel et al. 2002; Ghez et al. 2004). The luminosity is thought 
to be due to accreting thermal winds from its neighboring cluster of 
massive stars (e.g., Melia 1992). One of the key questions that has 
attracted much attention 
recently is why Sgr A* is so dim. The luminosity of Sgr~A* in each 
wavelength band is known to be about eight orders of magnitudes lower than 
the Eddington luminosity, prompting a number of theoretical models to 
explain its faint emission (e.g., Melia \& Falcke 2001; Yuan, Quataert \& 
Narayan 2003; Goldston, Quataert \& Tgumenshchev 2005;  Liu  \& 
Melia 2001; Liu, Melia \& Petrosian 2006). Theoretical models for Sgr A*'s 
spectrum require a very low 
efficiency for production of X-rays, considering the substantial mass 
inflow predicted from gas at the Galactic center.  Sgr A*'s multi-wavelength 
spectrum was fairly successfully modeled as a two-temperature 
radiatively inefficient flow which is advected across the event horizon 
before surrendering its luminosity--an ADAF (Narayan et al. 1998).  More 
recently, detection of radio and sub-millimeter polarization from Sgr A* 
(Bower et al. 2003; Marrone et al. 2006) has allowed 
estimates of the 
integrated electron 
density  in the accretion region from  $\sim$10 to  1000 Schwarzschild 
radii.  These estimates
indicate much lower electron densities  
than that predicted by the ADAF model, implying that most 
of the material is not reaching the central black hole.  Theoretical work 
(e.g., Yuan et al. 2003) supports this picture, but leaves unanswered the 
question of whether infalling material finds itself in a convective flow 
(Narayan et al. 2002), a (magnetically-driven) low-velocity outflow 
(Blandford \& Begelman 1999, Igumenshchev et al. 2003), or a fast jet 
(Yuan, Markoff \& Falcke 2002). 
By studying the flare emission from Sgr A*, we can potentially address
these issues. 

Recent multi-wavelength observations of Sgr A* show that near-IR flare 
activity is central
to the activity in X-ray and sub-mm wavelengths (e.g., Eckart et al. 2004, 
2006; Gillessen et al. 2006; Yusef-Zadeh et al. 2006). In particular, 
the
near-IR synchrotron emission is produced by a transient population of
$\sim$GeV electrons in a $\sim $10\,G magnetic field of size $\sim
10R_s$. Based on the measurements of the duration of flares that have
been detected at near-IR and sub-millimeter wavelengths, arguments have 
been made
that cooling could be due to adiabatic expansion, with the implication
that flare activity might be associated with  an outflow (Yusef-Zadeh et al. 
2006).
This interpretation stems from the fact that the synchrotron lifetime of
particles producing 850$\mu$m emission is about 12 hours, which is much
longer than the 20 to 40-min time scale for the GeV particles responsible 
for  
the near-IR emission. 
In order to test the expanding  model of flares, we carried 
out simultaneous radio continuum observations at 43 and 22 GHz  
to
search for a time delay in the peak frequency emission. 
The measurements presented here concentrate on 
the short term variability of Sgr A* whereas 
previous measurements 
at centimeter wavelengths have been concerned  with 
the long term variability of Sgr A* 
(Zhao et al. 2003; Herrnstein et al.\ 2004). 
Recent study of the long term variability  of Sgr A* at 
these wavelengths suggests that interstellar scattering is 
likely to be responsible for the flux density variation of 
Sgr A* (Macquart \& Bower 2005). Another important 
aspect of the present observations is that 
multiple frequency observations of Sgr A* have been carried 
out simultaneously for the first time 
using the fast-switching technique  at high frequencies, 

Here, we report a  weak flare that lasts for about two hours 
at 43 GHz ($\lambda$=7mm) and 22 GHz ($\lambda$=13mm). The 
cross-correlation plot  of 
the  radio data  shows a time delay which is accounted for in terms of  
a phenomenological  model of a uniformly expanding
plasma blob from Sgr A*. 

\section{Observations and Data Reductions} 

We used the Very Large Array (VLA) of the National 
Radio Astronomy Observatory\footnote{The National Radio Astronomy 
Observatory is a facility of the National Science Foundation, operated 
under a cooperative agreement by Associated Universities, Inc.} to observe 
Sgr A* simultaneously at 43 and 22 GHz in the BnA array 
configuration 
on Feb. 10 and 11, 2005 for 4 hours each.  The second day of the observing 
was washed out due to rain. The fast switching mode was used to quickly 
alternate between Sgr A* (4 minute) and a nearby (8\ddeg6 from Sgr A*) 
source 1820-254 (1 minute) at each frequency.  This allowed us to apply 
a complex gain calibration determined for the calibrators frequently 
throughout the run.  The calibration of the phases were done at high 
temporal resolution (every 10 seconds). 3C286 was used as a primary gain 
calibrator and NRAO 530 was observed every 15-30 minutes as a calibration 
check and in order to calibrate 1820-254.  The gains determined for
the calibrators were applied to Sgr A*, 
which was subsequently phase self-calibrated using the bright ($\sim$ 2 
Jy) point source itself for visibility spaces beyond 100 kilo$\lambda$.  An 
additional phase self-calibration was applied using the best clean 
component model, following a point-source self-calibration on Sgr A*. This 
allowed us to calibrate the phases for baselines longer than 20
kilo$\lambda$.  In order to compare the variability of Sgr A* against
that of a calibrator, we applied the derived complex gains from
NRAO 530 to 1820-254 and carried out an additional point-source phase
self-calibration. 

In order to examine if there is a time delay in the light curve of Sgr A* 
between 43 and 22 GHz data, we used the z-transformed 
discrete correlation function (ZDCF) (Alexander, 1997).  The ZDCF is an 
improved  
solution to the problem of investigating correlation in unevenly sampled 
light curves. The standard solutions are interpolation of the existing 
light curve, which is considered to be unreliable when power exists on 
smaller time scales than the gaps, and binning the data using discrete 
correlation functions (DCFs) (e.g. Edelson \& Krolik 1988). In DCFs, all 
pairs of points are ordered according to their time differences, binned, 
and compared.  The ZDCF improves upon standard DCFs by the application of 
the z-transform, which brings the skewed distribution of the DCF closer to 
a normal distribution.  The ZDCF chooses the binning of the data points to 
give equal populations in each bin, with a requirement that there be $>$11 
points in each bin.

\section{Results}

The light curves corresponding to 43 and 22 GHz data for Sgr A* 
and the nearby calibrator 1820-254 are shown in the top panels of
Figures 1 and 2, respectively. 
The fluxes presented in the light curves were derived by fitting a
point source at the phase center in the visibility plane using
spacings beyond 100 kilo-$\lambda$.  The measurements of the fluxes
from this technique were taken every 60 seconds.
The light curve of Sgr A* shows an increase of flux at a 
level of 7\% and 4.5\% at 43 and 22 GHz, respectively, when compared to 
its  quiescent flux,  whereas the  flux of the calibrator remains 
constant. For Sgr A*, the rise and fall time scale of 1.5 -- 2 hours 
noted in these measurements is similar to the flare time scale seen in 
sub-millimeter 
as well as millimeter wavelengths (Yusef-Zadeh et al. 2006; Mauerhan et al. 
2005; Eckart et al. 2006).  The variability analysis of several 
observations of Sgr A* at 43 GHz has also shown a 2-4 hour typical time 
scale (Roberts et al. 2006).  

The results of the cross correlation analysis of Sgr A* and 1820-254
are shown in the bottom panels of Figures 1 and 2, respectively. 
The cross-correlation peak of Sgr A*, as shown in Figure 1, is seen
at negative lag time, implying that the 22 GHz  flare can
not be leading the flare emission at 43 GHz. No peak is seen in the
cross-correlation plot of 1820-254 whereas Sgr A* shows the
appearance of two peaks with a delay of $\sim$ 20 and 50 minutes between 
43 and 22 GHz for
Sgr A*.  Due to the fact that the cross-correlation peaks are  broad,
we can not rule out the possibility of zero time delay. However, we
believe it is unlikely, given that both  peaks are  clearly not centered
at zero time delay. 

The spectral index measurements of the flaring and the quiescent
phase of Sgr A* have also been made. We begin by subtracting a constant flux
density of S$_{\nu}$ = 1.63 and 1.11 Jy from the combined flux
density of the quiescent and flare emission at 
43 and 22 GHz, respectively. The average spectral index
(S$_{\nu}\propto\nu^{\alpha}$) of the flare is
estimated to be $\alpha = 1.18 \pm$ 0.017, which 
is steeper by a factor of two than that of the
quiescent emission, (0.58 $\pm$ 0.004).  The spectral index of
the peak emission is 1.56 assuming a maximum flux density of the
flare S$_{\nu}^m$= 148 and 52 mJy at 
43  and 22 GHz,
respectively.   Since the measurements of the 
light curves of the flare at 
43  and 22 GHz
are interleaved throughout  the observing period, 
there is some uncertainty as to the
exact value of the peak flux density at 43 GHz,  considering that there 
is  no 
data around the time of the peak flux. Additional uncertainty in 
determining  the spectral index of the flare emission comes from  a 
simple average of
the parts of the light curve that was taken 
 on either side of the flare.  
Since the
flare started near the beginning of the run and the fact that the
quiescent emission appears to have structure (i.e., not completely
constant), a simple average is unlikely to  be  correct.   It is 
difficult to 
quantify the uncertainty which  could affect the peak flux of 
the flare emission. This is the first time that 
such  a flare emission has been seen simultaneously at 22 and 43 GHz 
using the fast-switching technique. We hope that future  
high-resolution VLA observations 
 address the true quiscent flux of Sgr A* at these high 
radio frequencies.   Recent 43 GHz measurements of Sgr A*
show a great deal of variability in the so-called quiescent flux of 
Sgr A* (Roberts et al. 2006).

 Given the  uncertainties described above, we measured the variation of
the spectral index as a function of time which 
shows  a steep spectral index of $\alpha\sim2.4$ at
the beginning of the flare before it flattens out towards the end of the
flare.  
The steep value of $\alpha$ at the beginning of the flare is
 due to the fact that the peak of emission at 43 GHz leads that at 22 GHz.
It is instructive to note that snapshot  observations at 
radio 
wavelengths   could give a misleading spectral index distribution when 
there is a time delay between the peak emission at different 
frequencies. 


\section{Discussion}
\subsection{The Plasmon Model}

The near-IR flaring of Sgr A* is generally thought to be due to optically-thin
synchrotron emission from a transient population of particles produced
within $\sim 10 $ Schwarzschild radii of the massive black hole (e.g.,
Genzel et al.  2003; Eckart et al.  2006; Gillessen et al. 2006; 
Yusef-Zadeh et al.  2006).
We have argued recently (Yusef-Zadeh et al.\ 2006) that the 
 sub-millimeter
flare observed simultaneously with a near-IR flare implies that the
emission is also optically thin at 350\,GHz ($\lambda$=857$\mu$m).  
 The $\sim 30$
minute duration of the flares at both frequencies strongly suggests that
the decline is not due to synchrotron cooling (estimated to be $\sim
20$ minutes and $\sim 12$ hours at 1.6 (188 THz) and 850 (350 GHz) \micron\ 
respectively) but
due to adiabatic cooling associated with expansion of the emitting
plasma.

The time delay we have observed between flaring at 43 and 22 GHz is
consistent with this picture: as the synchrotron optical depth
$\propto \nu^{-2.5}$, the emission at these frequencies is initially
optically thick.  The intensity grows as the blob expands, then peaks
and declines at each frequency once the blob becomes optically thin.
This first occurs at 43 GHz, and then at 23 GHz about 30--60 minutes
later.

To consider this quantitatively, we apply the plasmon model of van der
Laan (1966).  In this model, flaring at a given frequency is produced
through the adiabatic expansion of an initially optically-thick blob
of synchrotron-emitting relativistic electrons.  The initial rise of
the flux is produced by the increase in the blob's surface area while
it still remains optically thick; the curve turns over once the blob
becomes optically thin because of the reduction in magnetic field, the
adiabatic cooling of the electrons, and the reduced column density as
the blob expands.  The magnetic and the particle energy density both
drop as $R(t)^{-4}$ where $R$ is the increasing radius of the plasmon.
The model predicts simultaneous flaring
and declining emission at high frequencies, which are optically thin,
and increasingly delayed flaring at successively lower frequencies
that are initially optically thick.

Following van der Laan (1966), the synchrotron flux from a homogeneous 
blob can be written as
\begin{equation}
    S_\nu(R) = S_{0} \left(\frac{\nu}{\nu_0}\right)^{2.5}
    \left(\frac{R}{R_0}\right)^3 \,
    \frac{1-\exp(-\tau)}{1-\exp(-\tau_0)}
    \label{eq:Snu}
\end{equation}
where $R(t)$ is its radius, $p$ is the index of the relativistic particle 
energy spectrum
($n(E)\propto E^{-p}$),  and the optical depth
\begin{equation}
    \tau = \tau_0 \left(\frac{\nu}{\nu_0}\right)^{-(p+4)/2}
    \left(\frac{R}{R_0}\right)^{-(2p+3)}
    \label{eq:tau}
\end{equation}
Here $\tau_0$ is the critical optical depth at the maximum of the 
light curve at any frequency, which satisfies 
\begin{equation}
    e^{\tau_0}-(2p/3+1)\tau_0 - 1 = 0
    \label{eq:tau0}
\end{equation}
and $\nu_0$ is the frequency at which this occurs when $R=R_0$. 
The equivalent expression to (3) in van der Laan (1966) has a different
coefficient of $\tau_0$ because his expression is for the optical depth
corresponding to the maximum in the spectrum at some instant
($\partial_\nu S(\nu,t)=0$) rather than the maximum in the light curve at
a particular frequency ($\partial_t S(\nu,t)=0$).
To obtain numerical values, we assume
that a typical near-IR flare with flux 1\,mJy at 1.6$\micron$
(Yusef-Zadeh et al.\ 2006) is produced by a population of relativistic
electrons between 10\,MeV and 3\,GeV in
equipartition with the magnetic field.  The population is confined to
a blob characterized by initial size $2R_0$ and an optical depth that
yields 1\,mJy at 1.6$\micron$.

The remaining parameters in this model are the spectral index of
particles ($p$) and the initial blob radius  ($R_0$).  We first show the 
light curves (Fig.\ 3) at 350, 96, 43 and
22 GHz for the choices $p=3$ and $R_0=3R_s$ (for a
$3.7\times10^6$ M$_\odot$ black hole), which give a reasonable match to
our 43 and 22 GHz data.  For this choice of $p$, the critical optical
depth is $\tau_0 = 1.9$ and the corresponding peak frequency is
initially $\nu_0\approx 130$ GHz ($\lambda$=2.3mm).  The horizontal axis 
shows the
expansion factor over the original blob size; a form of $R(t)$ must be
adopted to map this to time, here we simply assume that $R\propto t$.
The blob is initially optically thin above 130 GHz, and so
the 350 GHz flux declines monotonically as the blob expands.  This is
consistent with our recent multi-wavelength observing campaign in
which apparently simultaneous flares were detected at 850 $\mu$m and
1.6$\mu$m, although we can  not rule out the possibility that the
sub-mm flare was delayed by about 100 minutes with respect to an
earlier near-IR flare (Yusef-Zadeh et al.  2006).  The peak at 96 GHz 
($\lambda$=3mm)
is attained rather quickly, as its initial optical depth is $\approx 6$
and drops as $R^{-9}$.  The amplitudes of the flares at 96 and
350 GHz are consistent with the intensity fluctuations observed at
these frequencies (Miyazaki, Tsutsumi \& Tsuboi 2004; Mauerhan et al.\ 
2005; Marrone et al.\ 2006);
unfortunately simultaneous observations at these frequencies are not
available at present.  The observed $\sim 30$ minute delay between the
peaks at 22 and 43 GHz sets the time scale for the light curves: the
plasmon must double its original size in $\sim 1$ hour.  We expect the
96 GHz flux to have peaked $\sim 30$ minutes before the 43 GHz
flare, and about 10 minutes after the near-IR and sub-millimeter
flares.

Now we turn to the dependence on the particle index $p$.  
Equations 1 and 2 imply that the frequency
at which the light curve is just peaking is
\begin{equation}
    \nu_p = \nu_0 \left(\frac{R}{R_0}\right)^{-(4p+6)/(p+4)}\,,
    \label{eq:nu_p}
\end{equation}
with peak flux
\begin{equation}
    S_p = S_0 \left(\frac{\nu_p}{\nu_0}\right)^{(7p+3)/(4p+6)}
        = \left(\frac{R}{R_0}\right)^{-(7p+3)/(p+4)}\,.
    \label{eq:S_p}
\end{equation}
The ``spectral index'' $\alpha$ of the peak fluxes ($S_p \propto \nu_p^\alpha$)
is only weakly dependent on the particle index $p$, ranging from 1.21
to 1.55 as $p$ runs from 2 to 8.  

To illustrate the above point, the peak fluxes attained at 350, 96, 43
and 22 GHz during the subsequent evolution of the $R_0=3R_s$ blob are
plotted as a function of the electron power-law index $p$ in Fig.\ 4.
The horizontal lines indicate nominal peak fluxes at the four
frequencies -- from our measurements at 22 and 43 GHz, and at 96 GHz
and at 350 GHz by Mauerhan et al.\ (2005) and by Marrone et al.\
(2006),  respectively.  Note that the nominal 96 and 350 GHz flux measurements 
have not 
been 
carried out simultaneously with those at  43 and 22 GHz. For 
the two lowest frequencies, the peak flux
always corresponds to the transition from optically thick to optically
thin.  In the mm and sub-mm, the initial blob is not optically thick
if $p\la 2.5$ and $p\la 3.7$ respectively, and the flux at those
frequencies then decays monotonically with time; hence the rapid
decline in peak flux at small values of $p$.

Several points are apparent from Fig.\ 4.  First, provided that the
blob is initially optically thick at the frequencies of interest,
there is a slow increase in spectral index of the peak fluxes as $p$
is increased, as dictated by equation 5.  Our observed
spectral index of 1.56 formally corresponds to $p=8.41$.  While this
is consistent with the spectral index range reported in near-IR
synchrotron flares (Eisenhower et al.  2005; Gillessen et al.  2006),
such a steep particle spectrum overproduces the flux in sub-mm flares
by a factor of 5.  In fact $p$ is very sensitive to $\alpha$ and so is
poorly determined by our observations.  For example, a 25\% increase in
the observed flux at 22 GHz would imply $p\approx 3$.  
 On the
other hand, the ratios of the peak fluxes at the three lower
frequencies \emph{are} fixed by this model \emph{and} match the
typical values that have been found (although the 96 GHz observations
do not strictly correspond to our particular flares).  Third, the
nominal peak fluxes at the three lower frequencies can be matched for
$p\approx 3$ or $p\approx 8$, but the sub-millimeter flux only matches 
for $p\approx
3$, when the initial blob is optically thin.  We therefore conclude
that $p\approx 3$ for our nominal fluxes.

The initial blob size $R_0$ and particle spectral index $p$ are
strongly constrained by the combination of measurements at frequencies
that are initially optically thin ($\nu>\nu_0$) and optically thick
($\nu<\nu_0$).  The initial flux at frequencies above $\nu_0$ is tied
to the assumed near-IR flux by the particle spectral index and do not
depend on the assumed source size.  The blob size does affect the peak
flux in the light curves of frequencies below $\nu_0$ by determining
the initial optical depth.  We have also plotted the peak fluxes
obtained for $R=5R_s$.  This increases the fluxes at 22, 43, and
96 GHz, and shifting the match to $p\approx 2.5$ and reducing the
350 GHz flux to $\sim 100$\,mJy.

The plasmon model, while phenomenological, is surprisingly
instructive.  It's clear that simultaneous monitoring of Sgr A* at
several frequencies between 10 and 100 GHz to determine the peak
fluxes and time delays will be able to test the model's applicability.
Adding simultaneous monitoring at two frequencies above
$\sim 200$ GHz, where the initial blob is optically thin, should allow
measurement of the flare parameters on a flare-by-flare basis.

Meanwhile, we can make some physical inferences based on the estimates
$p\approx 3$ and $R\approx 4R_s$.  The observed delay, $\sim 30$
minutes between 22 and 43 GHz, implies that the plasmon expands at a
rate $R_0$/hr, or $\approx 0.02\,c$.  The equipartition field strength
is $\approx 22$\,G, and the number density of relativistic electrons
between 10\,MeV and 3\,GeV is $\approx 6\times 10^5$\,cm$^{-3}$.
Adding an equal number of protons to preserve charge neutrality
implies a minimum blob mass $\sim 4\times 10^{19}$\,g.  In the absence
of significant external pressure the natural expansion speed is the
internal sound or Alfv\'en speed, $\sim 0.2c$.  The fact that it
expands more slowly implies either that it is confined by external
pressure or that its density is a hundred times greater than the estimate
above because of the presence of a thermal gas component.  The mass of
the blob would then be $\sim 4\times10^{21}$\,g.  If this component is
in energy equipartition with the field, its temperature would be $\sim 1\times
10^9$\,K. 

If the near-IR flares occur about once per hour and the blobs escape the 
system (though this appears highly unlikely given the slow inferred 
expansion speed) the mass loss rate, including the thermal component, 
would be $\sim 2\times10^{-8}$\,M$_\odot$\,yr$^{-1}$. This mass-loss rate 
is roughly similar or less than the mass accretion rate 
estimated  from  
the integrated electron density in the 
innermost $\sim$10 to 1000 Schwarzschild radii of the accretion disk 
imposed by the rotation measure measurements (Marrone et al, 2006). The 
combined accretion and mass-loss rate from Sgr A* is still less than the 
Bondi-Hoyle accretion rate of $\sim10^{-6}$\,M$_\odot$\,yr$^{-1}$ 
estimated from X-ray measurements (Baganoff et al. 2003). Given the 
0.5$''$ spatial resolution of {\it Chandra}, this suggests that the X-ray 
measurements of Sgr A* could be contaminated by thermal X-ray flux from 
the cluster of massive S stars distributed within 0.5$''$ of Sgr~A*. The 
spectral properties of this cluster of stars show massive stars but are 
generally consistent with normal main sequence O8V/B0V to B9V stars 
(Eisenhauer et al. 2005; Baganoff et al. 2003; Nayashkin \&
Sunyaev 2005). Early B stars tend to contribute between 10$^30$ 
and 10$^31$ ergs s$^{-1}$ per star in the 2-8 keV band (Feigelson et al. 
(2002).  
Ten of these early B-type stars combined with potentially X-ray emitting 
low-mass stars could contribute a 
substantial  portion of the soft diffuse emission, thus lowering the estimate 
of the Bondi rate. 
Alternatively, we speculate that the mass-loss rate from the 
young cluster  near Sgr A* is over-estimated or that the 
accreting thermal winds from the neighbouring cluster is obstructed as 
they approach Sgr A*.



\subsection{Alternative  Models}

There is a great deal of recent theoretical studies shedding light on 
other alternative models to explain the flaring activity of Sgr A*. The 
flare  emission has  also been  described by a jet model and an orbiting 
hot 
spot model (e.g., Eckart et al. 2006; Broderick \& Loeb 2006). However, we 
believe it is not necessary to discuss  these alternative models in 
detail as there is 
is 
no  evidence that Sgr A* showing  either a jet or a disk.  Given that the 
orbiting  hot spot model predicts no time delay in flare emission 
(see Figure 11 of Broderick and Loeb 2006),  we 
briefly view 
the expanding blob in the context of a jet model.

It is also possible that the expanding blob is confined by a jet that can 
account for the overproduction of sub-millimeter flare emission when the 
particle spectrum is steep. In the context of a jet model, one can 
consider the decreasing peak frequency in terms of an inhomogeneous jet 
model in which the magnetic field density and particle energy density 
decrease as powerlaws in distance with indices m and n, respectively 
(e.g., Konigl 1981).  The self-absorption frequency, then, is proportional 
to $r^{-k}$, where $k=[(2 + p) m + 2n -2]/(4+p)$.  We consider two 
distinct cases, where magnetic and particle energy density losses are 
alternatively dominant.  In both cases, we assume $p=3$ and an outflow 
that originates at 10 Schwarzschild radii with velocity $v \sim c$.  In 
the first case, we assume $m=1$ and $n=0$ and find $k=3/7$.  The time for 
the peak frequency to decrease by a factor of two is $t_2 \sim 1000$ sec.
 In the second case, we assume $m=0$ and $n=2$ and find $k=2/7$ and $t_2 
\sim 2000$ sec. These values are similar to the time delay observed 
between 43 and 22 GHz peak emission.  Thus, relativistic outflow in an 
inhomogeneous jet is another  possibility that can  account for the observed 
time 
delay in radio wavelengths.

Given that the overall characteristics of flare emission from Sgr~A*  
are  similar to those of 
X-ray binaries and microquasars, the jet model has been successfully applied
to a number of sources such as GRS 1915+105 (e.g., Mirabel et al. 1998; 
Fender \& Pooley 1998; 
Fender \& Belloni 2004) and  
Cyg X-3 where the frequency dependent of 
the peak flux density is
in the optically thick domain due to synchrotron self-absorption or
free-free absorption (Marti et al.  1992).  The jet model of Sgr A*
has also been explored in other contexts to explain the broad band 
spectrum of its   quiescent 
flux (e.g., Falcke \& Markoff 2000), 
its  near-IR
variability (Eckart et al.  2004; 2006), 
as well as the linear and circular polarization measurements (e.g., 
Falcke \& Beckert 2002).  

Additional circumstantial evidence that 
there may be collimated outflow from Sgr A* comes from the morphology of
ionized gas distributed outside the inner 0.5$''$ of 
 of Sgr A* 
(Yusef-Zadeh, Morris \& Ekers 1990; Zhao et al. 1991).
A   chain 
of blob-like structure was  noted best  to link to Sgr A* by a ridge of 
emission at 8 GHz 
(Wardle \& Yusef-Zadeh 1992).
These 
blobs are 
estimated to have a mass of 10$^{-2}$ \msol\ each with a size of about 
0.5  to 1$''$.  
 Although these 
blobs were considered to be formed as a result of the focusing of 
the IRS 16 cluster wind by the gravitational potential of Sgr A* (Wardle 
\& Yusef-Zadeh 1992), it is 
possible 
to consider the ridge of blob-like emission 
resulting directly from  a jet associated with   Sgr A*.   
Future proper motion measurements of the blobs as well as 
modeling of flaring jet 
emission from Sgr A* 
should be explored in more detail, in spite of a lack of direct 
evidence  for
collimated outflow from Sgr A*.

\section{Conclusions}

Radio continuum measurements using a fast switching technique between 
43 and 22 GHz frequencies show a weak flare with a duration of 
about two hours and time delay between the peak frequency emission. 
The adiabatic expansion of a uniform spherical synchrotron-emitting plasma blob 
based on 
the model  by van der Laan (1966)
can explain the overall characteristics of the observed flare
emission in multiple wavelengths.  In particular, the plasmon model is 
consistent with the time delay observed  between the peak 43 GHz and 22 GHz 
flare emission, the flux values at optically thin radio frequencies 
and at higher optically thin frequencies. 
Future simultaneous measurements at optically thin  and thick 
frequencies  should test this model.

\begin{figure}
\includegraphics[scale=0.8,angle=0]{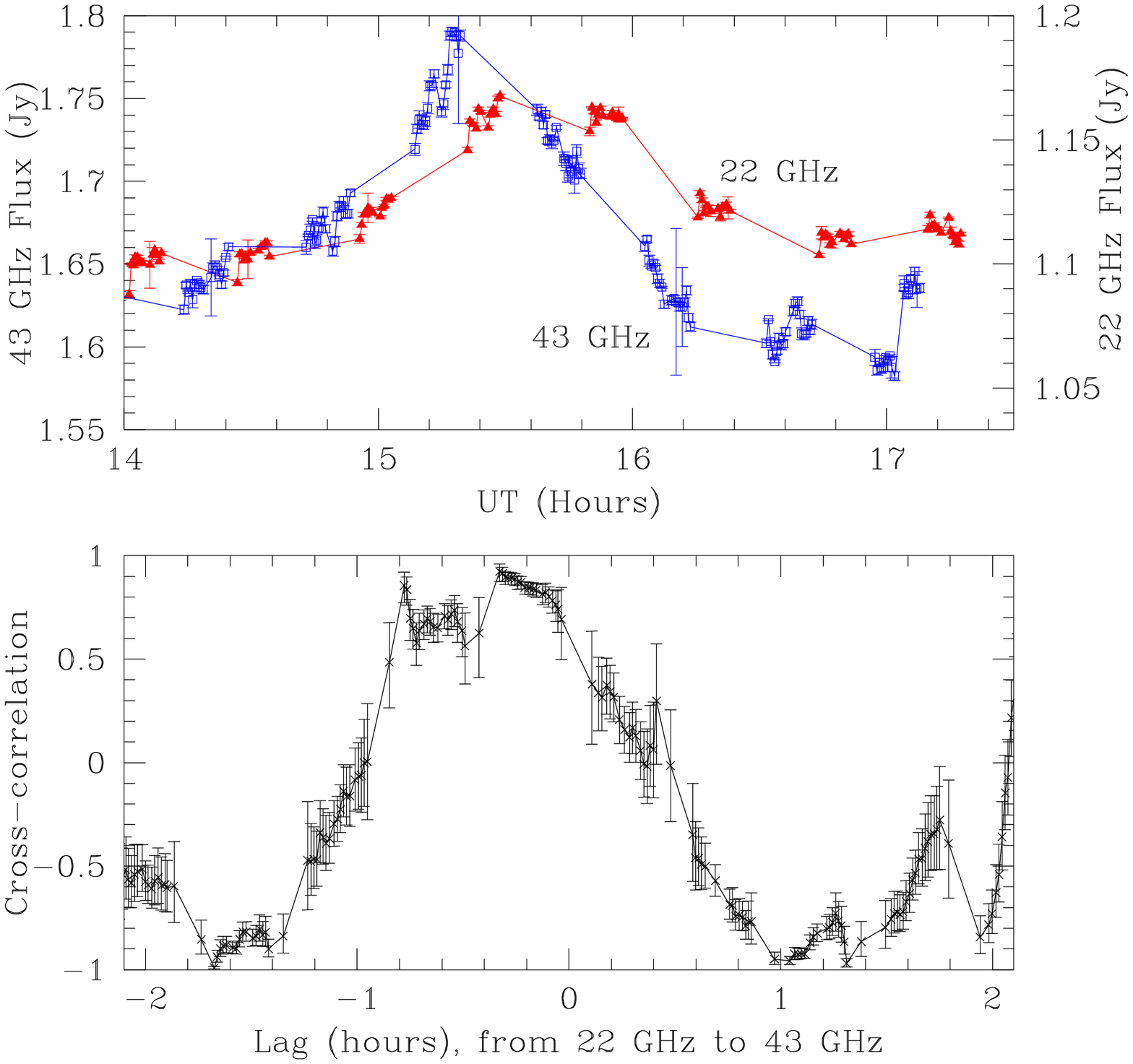}
\caption{\footnotesize (\textsl{Top}) The light curve
of Sgr A* flaring at 43  and 22 GHz with a sampling time of 30 seconds.
(\textsl{Bottom}) The cross-correlation amplitude as a function 
of lag time. 
}
\end{figure}

\begin{figure}
\includegraphics[scale=0.8,angle=0]{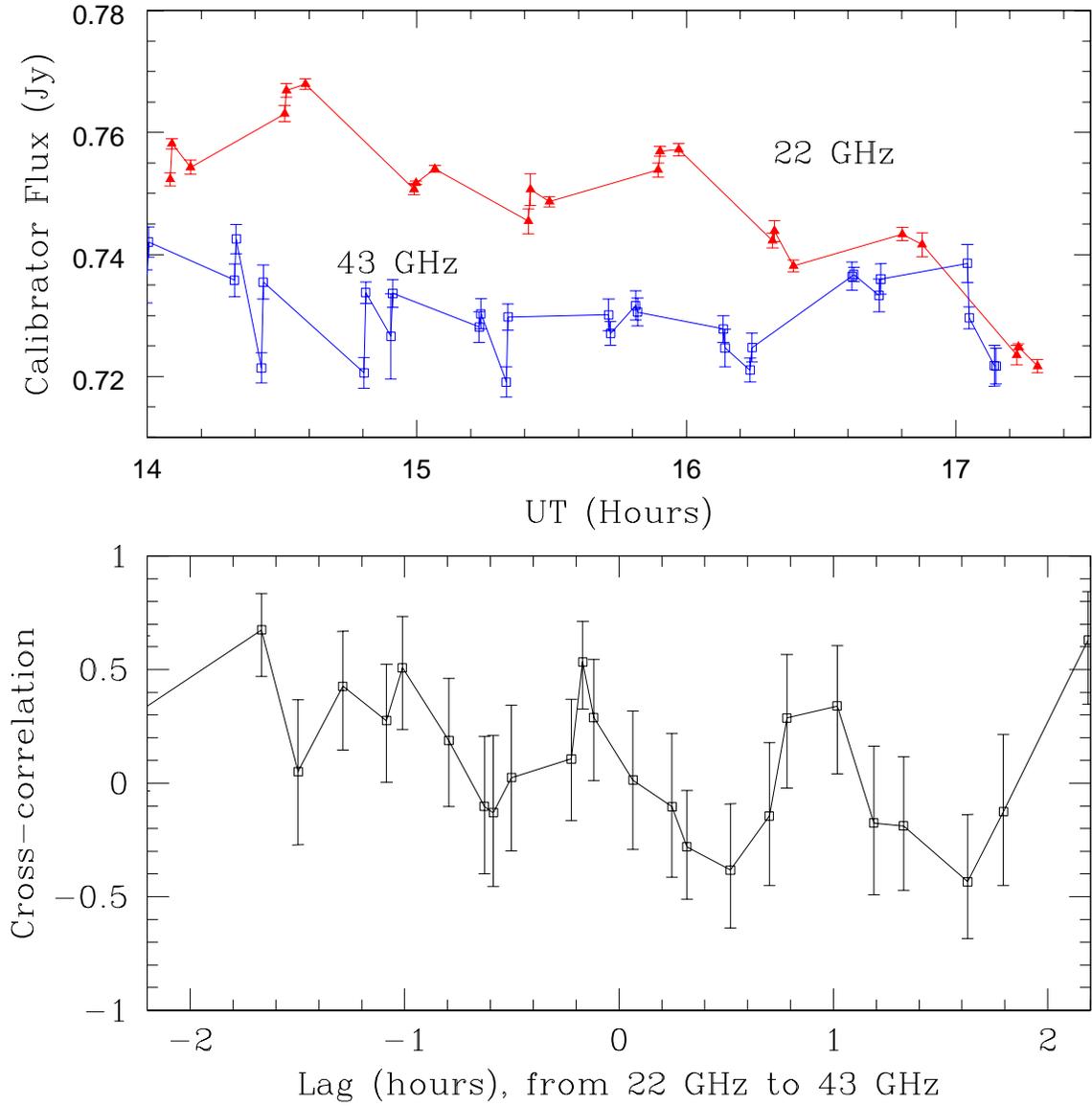}
\caption{\footnotesize (\textsl{Top}) The light curve
of 1820-254 showing no variability at 43  (blue) and 22 GHz (red).  
(\textsl{Bottom}) The cross-correlation amplitude as a function 
of lag time. The sampling time is 30 seconds. 
}
\end{figure}

\begin{figure}
\begin{center}
    \includegraphics[scale=1.0,angle=0]{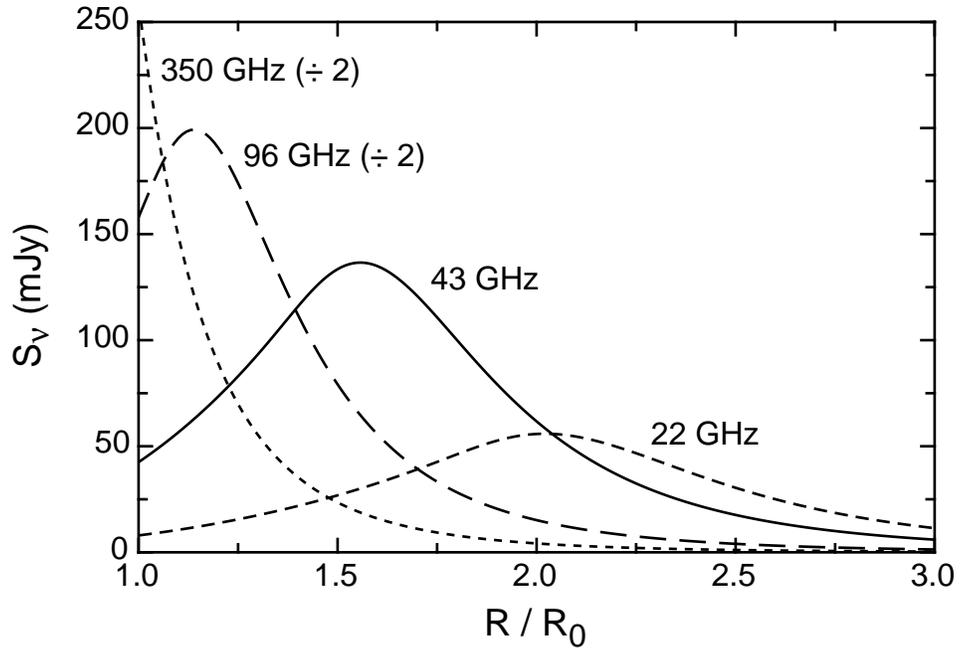}
\end{center}
\caption{\footnotesize Synchrotron light curves at four different
frequencies for an expanding blob of plasma with an $E^{-3}$ electron
spectrum in equipartition with its magnetic field.  The blob is
assumed to have an initial radius $R_0=4R_s$ and near-IR flux of
1\,mJy (see text).}
\end{figure}

\begin{figure}
\begin{center}
    \includegraphics[scale=1.0,angle=-90]{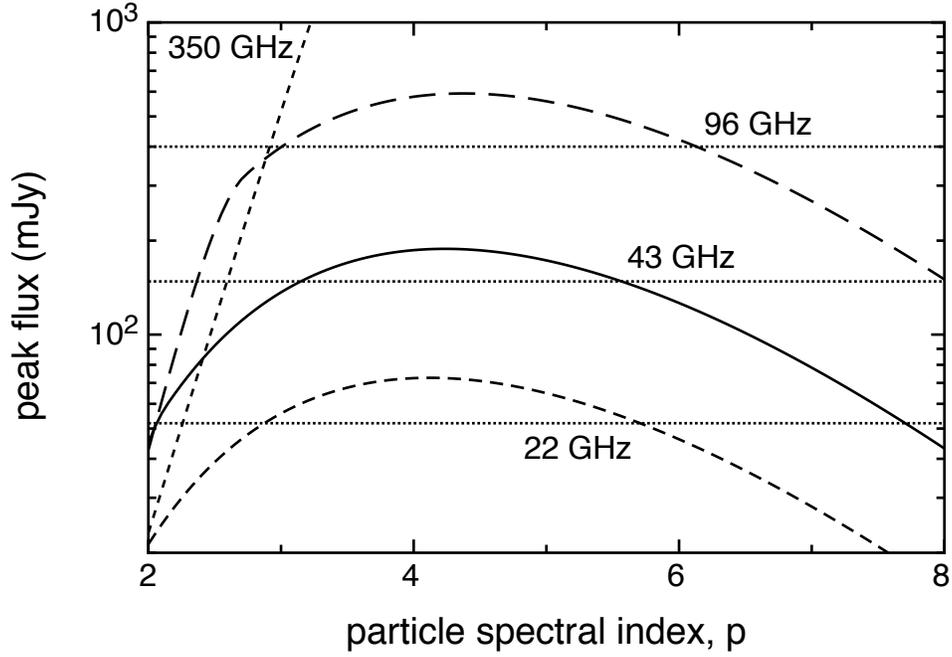}
\end{center}
\caption{\footnotesize Peak flux in the light curves from the blob at
four different frequencies as a function of assumed particle spectral
index $n(E)\propto E^{-p}$ (see text).  As in Fig.\ 3, the blob is
assumed to have an initial radius $R_0=4R_s$ and near-IR flux of
1\,mJy.  The dotted horizontal lines indicate the observed peak fluxes
of 148, 52 and 400 mJy
at 22 and 43 GHz and a typical flare amplitude at 96 GHz (Mauerhan et 
al. 2005), respectively. The typical flux of a flare at 
 350 GHz is assumed to be  
$\sim$0.5 Jy. The only simultaneous flux measurements are taken at 43 and 22 
GHz.  }
\end{figure}


\begin{references}
\noindent
Alexander, T. 1997, in Astronomical Time Series, ed. D. Maoz, A. Sternberg 
\& E. Leibowitz (Dordrecht:Kluwer), 163

\noindent 
Balick,  A. \& Brown, R.L.  1974, ApJ, 194, 265


\noindent
Baganoff, F.K., Maeda, Y., Morris, M., Bautz, M.W., Brandt, W.N. et al. 
2003, ApJ, 591, 891

\noindent
Blandford, R. D. \&  Begelman, M.  C. 1999, MNRAS, 
303, L1

\noindent
Bower, G.C., Wright, M.C., Heino, F. \& Backer, D.C. 2003, ApJ, 588, 331

\noindent
Broderick, A.E. \& Loeb, A. 2006, MNRAS, 367, 905

\noindent
Eckart, A.,  Baganoff, F. K.,  Morris, M.,  Bautz, M.W., Brandt, W.N.
et al.\  2004, A\&A, 427, 1

\noindent
Eckart, A., Baganoff, F.K., Schoedel, R., Morris, M., Genzel, R., Bower, 
G.C. et al. 2006, A\&A, (in press) (astro-ph/0512440)	

\noindent
Edelson, R.A.  \& KroliK, J.H.  1988, ApJ, 333, 646 

\noindent
Eisenhauer, F., Genzel, R., Alexander, T., Abuter, R., Paumard,
T.,  Ott, T.,  Gilbert, A.,  Gillessen, S.,  Horrobin, M.,  Trippe, S. et
al. 2005, ApJ, 628, 246

\noindent
Beckert, T. \& Falcke, H. 2002, A\&A, 388, 1106

\noindent
Broderick, A.E. \& Loeb, A. 2006, MNRA, 367, 905

\noindent
Falcke, H. \& Markoff, S. 2000, A\&A, 362, 113

\noindent
Fender, R. \& Belloni, T. 2004, ARAA, 42, 317 

\noindent
Fender, R.P. \& Pooley, G.G. 1998, MNRAS, 300, 573

\noindent
Feigelson, E. D.,  Broos, P.,  Gaffney, J. A.,  Garmire, G.  
Hillenbrand, L.  A. et al. 2002, ApJ, 574, 258


\noindent
Genzel, R.,  Sch\"odel, R., Ott, T., Eckart, A.,
Alexander, T.,  Lacombe, F., Rouan, D.,  Aschenbach, B. et al. 2003,
Nature,  425, 6961, 934

\noindent
Ghez, A. M.,  Wright, S. A.,  Matthews, K., Thompson, D.,
Le Mignant, D.,  Tanner, A., Hornstein, S. D.,  Morris, M.,
Becklin, E. E. \& Soifer, B. T. 2004, ApJ, 601, L159

\noindent
Ghez, A. M.,  Hornstein, S.D., Lu, J.,  Bouchez,
A.,  Le Mignant, D. et al. 2005, ApJ, 635, 1087

\noindent
Gillessen, S., Eisenhauer, F., Quataert, E., Genzel, R., Paumard, T.,
Trippe, Ott, T. et al. 2006, ApJ, in press (astro-ph/0511302)

\noindent
Goldston, J., Quataert, E. \& Tgumenschchev, I.V. 2005, ApJ,
621, 785

\noindent
Herrnstein, R.M.,  Zhao, J.-H., Bower, G.C. \&
Goss,  W. M. 2004, AJ, 127, 3399

\noindent
Igumenshchev, I.V.,  Narayan, R. \& Abramowicz, M. A., 2003, ApJ, 592, 
1042

\noindent
K\"onigl, A. 1981, ApJ, 243, 700

\noindent
Liu, S. \& Melia, F. 2001, ApJ, 561, L77

\noindent
Liu, S., Melia, F. \& Petrosian, V. 2006,
ApJ, 636, 798

\noindent
Macquart, J.-P. \& Bower, G.C., 2006, ApJ, in press

\noindent
Marrone, D.P., Moran, J., Zhao, J.-H. \& Rao, R. 2006, ApJ, in press 
(astro-ph/0511653)

\noindent
Marti, J., Paredes, J.M. \& Estalella, R. 1992, A\&A, 258, 309

\noindent
Mauerhan, J.C., Morris, M., Walter, F.  \&  Baganoff, F.   2005, ApJ, 623,
L25

\noindent
Melia, F. 1992, ApJ, 387, L25

\noindent
Melia, F., \& Falcke, H. 2001, ARAA, 39, 309



\noindent
Mirabel, I.F., Dhawan, V., Chaty, S., Rodriguez, L.F., Marti, J., 
Robinson, C.R. et al. 1998, A\&A, 330, L9


\noindent
Miyazaki, A., Tsutsumi, T. \& Tsuboi, M. 2004, ApJ, 611, L97 

\noindent
Narayan, R., Mahadevan, R., Grindlay, J.E., Popham, R.G \& Gammie, C. 	
1998, ApJ, 492, 554

\noindent
Narayan, R.,  Quataert, E., Igumenshchev, I. V. \& Abramowicz, M. A. 
2002, ApJ, 577, 29

\noindent
Nayakshin, S. \& Sunyaev, R. 2005, MNRAS, 364, L23

\noindent
Roberts et al. 2006, in preparation

\noindent 
Sch\"odel, R., Ott, T., Genzel, R.,  Hofmann, R., 
Lehnert, M., Eckart, A.,  Mouawad, N.,  Alexander, T.,  Reid, M. J. \& Lenzen
R. 2002,  Nature, 419, 694

\noindent
van der  Laan, H.  1966, Nature, 5054, 1131

\noindent
Wardle. M. \& Yusef-Zadeh, F. 1992, Nature, 357, 308

\noindent
Yuan, F., Markoff, S.,  \& Falcke, H. 2002, A\&A, 854, 383

\noindent
Yuan, F., Quataert, E. \& Narayan, R. 2003,  ApJ, 598, 301

\noindent
Yusef-Zadeh, M., Morris, M. \& Eckers, R. 1992, Nature, 348, 45

\noindent
Yusef-Zadeh, F.,  Bushouse, H.,  Dowell, C. D.,  Wardle, M., 
Roberts, D., 
Heinke, C. O.,  Bower, G. C. et al. 2006, ApJ, in press


\noindent
Zhao, J.-H.,  Goss, W. M., Lo, K. Y. \&  Ekers, R. D.
1991, Nature, 354, 46

\noindent
Zhao, J.-H., Young, K.H., Hernstein, R.M. et al. 2003,
ApJ, 586, L29


\end{references}
\end{document}